\documentclass{INTERSPEECH2023}

\usepackage{times}
\usepackage{latexsym}
\usepackage{multirow}
\usepackage{times}
\usepackage{latexsym}
\usepackage{multirow}
\usepackage{graphicx}
\usepackage{multirow, boldline}
\usepackage{subfigure}
\usepackage{xcolor}

\usepackage{makecell}
\usepackage{enumitem}

\usepackage{xcolor}
\usepackage{hyperref}

\usepackage{mathtools}

\makeatletter
\newcommand*\concat
  {\mathbin{\mathmakebox[\widthof{${+}\m@th$}]{+\hskip-1emplus1fil+}}}

\makeatother

 \newcommand\citep{\cite}
 
\usepackage{soulutf8}
\usepackage{todonotes}
\usepackage{placeins}
\usepackage{tikz}
\usepackage{svg}

\interspeechcameraready
 
\title{{ASR data augmentation in low-resource settings using cross-lingual multi-speaker TTS and cross-lingual voice conversion}}

\name{Edresson Casanova$^{1,2}$, Christopher Shulby$^3$,  Alexander Korolev$^4$, Arnaldo Candido Junior$^5$, Anderson da Silva Soares$^6$, Sandra Aluísio$^2$, Moacir Antonelli Ponti$^{2,7}$}

\address{ \normalsize
    $^1$ Coqui, Germany; $^2$ Instituto de Ciências Matemáticas e de Computação, Universidade de S\~ao Paulo, Brazil; \\
    $^3$ QuintoAndar, Portugal; $^4$ Darmstadt University of Applied Sciences, Germany; $^5$ São Paulo State University, Brazil; \\ $^6$ Federal University of Goiás, Brazil; $^7$ Mercado Livre, Brazil.
  }
  
\email{edresson@coqui.ai}

\begin{document}

\maketitle

\begin{abstract}
    {
    We explore cross-lingual multi-speaker speech synthesis and cross-lingual voice conversion applied to data augmentation for automatic speech recognition (ASR) systems in low/medium-resource scenarios. Through extensive experiments, we show that our approach permits the application of speech synthesis and voice conversion to improve ASR systems using only one target-language speaker during model training. We {also managed to close the gap between ASR models trained with synthesized versus human speech compared to other works that use many speakers.} Finally, we show that it is possible to obtain promising ASR training results with our data augmentation method using only a single real speaker in a target language.}
\end{abstract}

\vspace{-0.2cm}
\section{Introduction}
\label{sec:intro}
\vspace{-0.1cm}

Text-to-Speech (TTS) systems have received a lot of attention in recent years due to the great advances in deep learning, which have allowed for massive use in applications such as virtual assistants. These advances have allowed TTS models to achieve naturalness similar to human speech \cite{tacotron2, valle2020flowtron, kim2020glow, kim2021conditional}. Still, most TTS systems are tailored for a single speaker, where many applications could benefit from new-speaker synthesis, i.e., not seen during training, employing only a few seconds of the target speech. This approach is called zero-shot multi-speaker TTS (ZS-TTS) \citep{jia2018transfer, choi2020attentron, yourtts}.


Advances in TTS have motivated works that exploit it as a way to improve Automatic Speech Recognition (ASR). Researchers have explored two different approaches. The first is parallel training of ASR and TTS models; in this approach the TTS and ASR systems can improve themselves, as in \cite{tjandra2018machine,tjandra2020machine}. The second is the use of a pre-trained TTS model to generate new ASR training data, such as \cite{li2018training}, \cite{rosenberg2019speech} and \cite{laptev2020you}.  In this work, we will focus on the latter approach.

{ 
Many studies that explore a pre-trained TTS model to generate ASR data used the LibriSpeech dataset \cite{librispeech} to train the ASR model. For the TTS model training, \cite{li2018training} used 3 speakers from the American English M-AILABS dataset \cite{solak2019m}, while \cite{rosenberg2019speech} and \cite{laptev2020you} trained the TTS model with more than 251 speakers from LibriSpeech.} In Table~\ref{tab:related-works} we report the Word Error Rate (WER) of the best experiment from the related works in the test-other subset of the LibriSpeech dataset. Although the studies contain both test-clean and test-other results, we focus on the results of the most difficult sub-set. Also, \cite{laptev2020you} reported results using an external language model (LM); however, for fairness, we omit this LM in our comparison. 

\begin{table}[h]
\centering
\caption{Related works comparison in the test-other subset.}
 \vspace{-0.2cm}
\label{tab:related-works}
\resizebox{0.45\textwidth}{!}{%
\begin{tabular}{c|c|c|c|c}
\hline
\multicolumn{1}{c|}{\textbf{Paper}  }                     & \multicolumn{1}{c|}{\textbf{TTS Model}} & \multicolumn{1}{c|}{\textbf{ASR Model}} & \multicolumn{1}{c|}{\textbf{Train data}} & \textbf{WER} \\ \hline
\multirow{3}{*}{\cite{li2018training}}      & \multirow{3}{*}{ \begin{tabular}[x]{@{}c@{}} Tacotron \\ GST \end{tabular}}           & \multirow{3}{*}{Wav2Letter}             & Human                                    & 16.21        \\ \cline{4-5} 
                                     &                                         &                                         & Synthesized                              & 81.78        \\ \cline{4-5} 
                                     &                                         &                                         & \begin{tabular}[x]{@{}c@{}}Hum + Synt    \end{tabular}                  & 15.47        \\ \hline
\multirow{3}{*}{\cite{rosenberg2019speech}} & \multirow{3}{*}{\begin{tabular}[x]{@{}c@{}}ZS-Tacotron  \\  + VAE \end{tabular}}        & \multirow{3}{*}{LAS}                    & Human                                    & 13.89        \\ \cline{4-5} 
                                     &                                         &                                         & Synthesized                              & 66.10        \\ \cline{4-5} 
                                     &                                         &                                         & \begin{tabular}[x]{@{}c@{}}Hum + Synt    \end{tabular}                         & 13.78        \\ \hline
\multirow{3}{*}{\cite{laptev2020you}}       & \multirow{3}{*}{\begin{tabular}[x]{@{}c@{}} GMVAE \\ Tacotron \end{tabular}}        & \multirow{3}{*}{LAS}                    & Human                                    & 14.10         \\ \cline{4-5} 
                                     &                                         &                                         & Synthesized                              & --           \\ \cline{4-5} 
                                     &                                         &                                         & \begin{tabular}[x]{@{}c@{}}Hum + Synt    \end{tabular}                        & 13.50          \\ \hline
\end{tabular}
}
 \vspace{-0.5cm}
\end{table}

The ASR models trained with synthesized speech combined with human speech achieved relative improvement\footnote{{We used relative improvement/difference metric to show the real improvement achieved by related works' approaches.}} of 4.56\%, 0.79\% and 4.25\% compared to the models trained with human speech alone, respectively, for \cite{li2018training}, \cite{rosenberg2019speech} and \cite{laptev2020you}. {A greater difference is observed between the model trained with only human speech and only synthesized speech, with a relative difference of 80.17\% and 78.98\%, respectively, for \cite{li2018training} and \cite{rosenberg2019speech}, which motivates further improvements and research.
}

{In parallel with our work, \cite{baas22_interspeech} explored cross-lingual voice conversion (VC) for ASR data augmentation in low-resource settings. They showed that when using a sensible amount of voice conversion data augmentation, ASR performance is improved in all low-resource languages explored.}


{Although previous work shows the potential for multi-speaker TTS models for ASR data augmentation, these models still require high-quality datasets with many speakers and hours of speech to converge \cite{laptev2020you}. Generally, such models are trained on English with big datasets such as LibriSpeech \cite{librispeech} and LibriTTS \cite{zen2019libritts}, which is not suitable for low-resource languages that do not have an open multi-speaker TTS dataset.}


{Although some multilingual multi-speaker datasets were released in recent years \cite{park2019css10, pratap2020mls, elizabeth2021multilingual}, they just attend a small number of languages and for many applications, even these may not be sufficient to build a competitive ASR system. In addition, creating a high-quality multi-speaker dataset is hard, because it requires the effort of multiple target-language speakers. It is especially hard for languages with small populations, where recruiting participants is difficult or in more extreme scenarios with languages that are almost extinct and have just a few speakers (e.g. indigenous languages). In a range of scenarios creating a high-quality multi-speaker dataset is not viable. }

{In light of this, an approach that applies TTS/VC for ASR data augmentation that requires just a medium/low-quality single-speaker dataset could make the application of this technology viable for languages that really need it, helping to preserve/protect nearly extinct languages, for example. The objective of this paper is to improve upon such issues and make it viable. Here, we seek to answer two questions: Is a TTS model trained with just one speaker in a given target language sufficient for ASR data augmentation? Is only one human speaker in the target language enough to get a reasonable ASR model via cross-lingual voice conversion and cross-lingual multi-speaker TTS ASR data augmentation? The contributions of this work are as follows:}
 

\begin{itemize}
 \vspace{-0.1cm}
    \item {A novel approach for ASR data augmentation that explores cross-lingual voice conversion and a cross-lingual multi-speaker TTS model. For TTS and voice conversion. We used the YourTTS model \cite{yourtts}, which was developed in a previous work to meet the requirements needed for this paper. Our novel approach requires just one speaker in the target language, making the application of this technology possible for low-resource languages;}

    \item {We are the first to combine  multi-speaker TTS and voice conversion for ASR data augmentation. In addition, we are the first to explore cross-lingual multi-speaker TTS and cross-lingual voice conversion using speakers of other languages to fill the lack of speakers for low-resource ASR model training.}

    \item {We are the first to explore ASR data augmentation via TTS with a very limited number of speakers. To do so, we emulate a scenario where the ASR model and the TTS model would need to be trained with just one real speaker (low-resource language scenario) on two target languages. Our novel approach improves WER from 64-74\% to 34-37\%, approximately a 33\% absolute improvement. Such results indicate the feasibility of applying this technology for low-resource languages.}

\end{itemize}

\vspace{-0.2cm}
\section{Audio datasets}\label{sec:method:base}
\vspace{-0.1cm}

We used 3 languages/training datasets for the TTS model: 

\textbf{English}: VCTK dataset \cite{veaux2016superseded}, containing 44 {hours} of speech from 109 speakers, sampled at 48KHz. We divided the VCTK dataset into training, development and test subsets following \cite{yourtts}. To further increase the number of training speakers, we used the subsets \textit{train-clean-100} and \textit{train-clean-360} from LibriTTS \cite{zen2019libritts}. {In the end, our TTS model was trained with approximately 298 hours from 1,248 English speakers.}

\textbf{Portuguese}: TTS-Portuguese Corpus \citep{casanova2020ttsportuguese}, a single-speaker male dataset in Brazilian Portuguese (pt-BR) containing approximately 10 hours, sampled at 48KHz. As the authors did not use a soundproof studio, the dataset contains some environmental noise. 
Following \cite{yourtts}, we resampled the audios to 16Khz and used FullSubNet \citep{Hao_2021} as a denoiser. For development, we randomly selected 500 samples, leaving the rest for training.


\textbf{Russian}: ru\_RU set of the M-AILABS dataset based on LibriVox, consisting of 46 {hours} from 1 female and 2 male speakers. We used samples only from the female speaker for diversity, since we already used a male for pt-BR. For development, we randomly selected 500 samples, leaving the rest for training.

For all TTS datasets, pre-processing was carried out to normalize volume and to remove long silences, following \cite{yourtts}. After pre-processing, the datasets contained 8.38 {hours} for pt-BR and 14.94 {hours} for ru-RU (Russian). 

For ASR training, we used Common Voice version 7.0 \cite{ardila2020common} for pt-BR and ru-RU. In all experiments, we used the default train, development and test partitions. For pt-BR, these sets have 14.52, 8.9 and 9.5 hours, respectively; and ru-RU has 25.95, 13.06 and 13.75 hours, in the same order. {
In addition, the speaker distribution for train, development and test partitions are 103, 238 and 1252 speakers for pt-BR; and ru-RU has 117, 210 and 1202 speakers.}

\vspace{-0.2cm}
\section{TTS model setup}\label{sec:method:yourtts}
\vspace{-0.1cm}

{In our previous work, we presented YourTTS model \cite{yourtts}, a multilingual zero-shot multi-speaker TTS model, which achieved  state-of-the-art (SOTA) results using only a single-speaker dataset in the target language. Although the focus of the model is on TTS it can also do zero-shot voice conversion. This model was developed to meet the requirements needed for this paper. In the original work, we trained YourTTS using English (LibriTTS and VCTK datasets), French (M-AILABS dataset) and pt-BR (TTS-Portuguese Corpus). The model was trained using only one male pt-BR speaker, but still produced strong results in zero-shot multi-speaker TTS and zero-shot voice conversion for pt-BR. Furthermore, it was able to produce female voices even without being trained on female voices, making it adequate for the objective of this study. }


{Here, we fine-tuned the YourTTS model in English, pt-BR and ru-RU. For this we used transfer learning from the original checkpoints made publicly available. The dataset in English and pt-BR were the same dataset used in {our previous work}, but we replaced the French M-AILABS dataset with a female speaker from the ru-RU M-AILABS dataset due to experiment requirements as explained in Section \ref{sec:method:generated_vs_human}.}
We trained the YourTTS model for 140k steps with the same parameters used in \cite{yourtts}. In summary, we trained YourTTS with 1,248 speakers in English, 1 male pt-BR speaker and 1 female ru-RU speaker. {After the training, this model checkpoint was used as TTS/VC model for all experiments in this paper.}

YourTTS can synthesize different audios for the same input sentence. During inference, the latent variable, predicted by the text encoder, is added with a random sample of the standard normal distribution multiplied by a temperature $T$. In this way, diversity can be controlled by the temperature $T$. As shown by \cite{kim2020glow}, the manipulation of $T$ allows for generating different pitches; for more details see \cite{kim2020glow, kim2021conditional, yourtts}. Furthermore, YourTTS is trained with the stochastic flow-based duration predictor proposed in \cite{kim2021conditional}, which can produce several different speech rhythms for the same sentence. To do so, during inference, a random sample of the standard normal distribution is multiplied by a temperature $T_{dp}$ and added to the latent variable before being inverted by the flow. In this way, it is possible to control the variety of rhythms with the temperature $T_{dp}$ \cite{kim2021conditional}. 
Finally, it is possible to control the speaking rates by multiplying the predicted durations by a positive scalar $L$, thus making the pronunciation faster when $L$ is smaller and slower when $L$ is bigger. 

 \vspace{-0.2cm}
\section{{Is a multi-speaker TTS model trained with just one speaker in the target languages enough for ASR data augmentation?}}\label{sec:method:generated_vs_human}
\vspace{-0.1cm}

{Previous works have shown a large gap between ASR models trained with human and synthesized speech \cite{rosenberg2019speech, laptev2020you}, where researchers have used a large number of speakers and hours in the target language during the TTS training. To apply this method to languages with low/medium resources, we need an approach that only requires a single speaker in the target language. In this section, we propose to employ  YourTTS trained with only one speaker in the target language to do data augmentation for ASR. {It is important to note that this TTS is also trained in English and it is exposed to embeddings from English speakers}. We follow related works comparing ASR models trained with synthesized and human speech, to verify if our approach can be used as data augmentation for ASR.}
 

{For fair comparison between human and synthesized speech, we employ YourTTS model to  generate a synthesized version of pt-BR and ru-RU Common Voice datasets. For each sentence in Common Voice, we generate its pronunciation for the same speaker, using the sentence's pronunciation as reference for speaker embedding extraction. So, we have used speaker embeddings from the target language's native speakers. 
The idea being that if the zero-shot multi-speaker TTS model is good enough, it will generate the same speaker's voice as in the original audio, additionally, the synthesized and human dataset will have the same vocabulary.}{ During the generation,  as explained previously in Section \ref{sec:method:yourtts}, diversity is achieved by randomly choosing $L$, $T$ and $T_{dp}$: for $L$, a value between 0.7 and 2, while for temperatures ($T$ and $T_{dp}$) a value between 0 and 0.667.}

{To increase the diversity in ASR training, in some experiments, we also explored three popular augmentation methods in speech processing -- additive noise, pitch shifting and room impulse response (RIR) simulation. For additive noise and RIR filters we have used the same approach and dataset from \cite{heo2020clova}. For pitch shift, we randomly chose a semitone from -4 to 4.  All augmentations are randomly selected with a 25\% probability of being chosen for each audio in every training step. For all methods, we used the implementations available in the Python Audiomentations\footnote{https://github.com/iver56/audiomentations} package. We will refer to it as audio augmentations (AA).}

As for the ASR model, we use Wav2vec 2.0 \cite{baevski2020wav2vec}, a large self-supervised model trained on the VoxPopuli dataset \cite{wang-etal-2021-voxpopuli}. We used the model checkpoint provided by the authors\footnote{https://huggingface.co/facebook/wav2vec2-large-100k-voxpopuli} which was trained on 100k hours of speech in the following 23 languages: Bulgarian (Bg), Czech (Cs), Croatian (Hr), Danish (Da), Dutch (Nl), English (En), Estonian (Et), Finnish (Fi), French (Fr), German (De), Greek (El), Hungarian (Hu), Italian (It), Latvian (Lv), Lithuanian (Lt), Maltese (Mt), Polish (Pl), Portuguese (Pt), Romanian (Ro), Slovak (Sk), Slovene (Sl), Spanish (Es) and Swedish (Sv). We chose Pt and Ru because Pt was used in the {self-supervised} pre-training and Ru was not, presenting realistic results for both scenarios and are from different language families. We carried out four experiments: 
\begin{itemize}[noitemsep]
    \item \textbf{Experiment 1:} ASR models trained in pt-BR and ru-RU with Common Voice using the standard training and development subsets. For pt-BR, the model was trained for 140 epochs and ru-RU for 100;
    \item \textbf{Experiment 1.1:} Transfer learning from experiment 1, {adding audio augmentations (AA) in the training}. In this experiment, the ASR is trained with half the number of epochs used in experiment 1.  
     \item \textbf{Experiment 2:}  Similar to experiment 1, but the model is trained using {the synthesized version of Common Voice} in pt-BR and ru-RU. For model training and development, we used synthesized speech.
     \item \textbf{Experiment 2.1:}  Transfer learning from experiment 2 plus the AA. The ASR is trained with half the number of epochs used in experiment 2.
\end{itemize}

To run the experiments, we use the HuggingFace Transformers framework\footnote{https://github.com/huggingface/transformers}. The models were trained with a NVIDIA TITAN RTX 24GB GPU using a batch size of 8 and gradient accumulation over 24 steps. We used the AdamW optimizer with linear learning rate warm-up from 0 to 3e-05 in the first 8 epochs and after using linear decay to zero. 
During training, the best checkpoint was chosen, using the loss in the development set and early stopping after the development loss had not improved for 10 consecutive epochs. The code used for all of the experiments, as well as model checkpoints are publicly available at: \url{https://github.com/Edresson/Wav2Vec-Wrapper}.


{Table \ref{tab:results-comp} presents the WER results for our experiments on the original pt-BR and ru-RU test subsets of Common Voice.}

\begin{table}[!ht]
\centering
 \vspace{-0.3cm}
\caption{Human and synthesized speech comparison (WER).}
 \vspace{-0.2cm}
\label{tab:results-comp}
\begin{tabular}{l|c|c}
\hline
\multicolumn{1}{c|}{\textbf{Exp.}} & \textbf{PT}    & \textbf{RU}    \\ \hline
1. Human Speech                      & 23.50 &  25.47\\ \hline
1.1 Human Speech + AA                 &  21.54&  22.27\\ \hline
2. Synthesized speech                    & 56.84 & 65.85 \\ \hline
2.1 Synthesized speech + AA                &  43.99 & 50.46 \\ \hline
\end{tabular}
\vspace{-0.2cm}
\end{table}
The model trained only with human speech (Experiment 1) reached a WER of 23.50\% and 25.47\%, respectively for pt-BR and ru-RU. The model trained only with synthesized speech (Experiment 2) reached a WER of 56.84\% and 65.85\%, for pt-BR and ru-RU.  {Therefore, without AA, the relative difference between the model trained with only human speech and synthesized speech is of 58.65\% and 61.32\% for those two languages.} 

As expected, fine-tuning the models with AA (Experiment 1.1 and Experiment 2.1) improved performance. The model trained with human speech only improved its result by 1.96\% and 3.20\% WER for pt-BR and ru-RU after the addition of AA. The model trained only with synthesized speech improved by 12.85\% and 15.39\% WER. Therefore, using AA benefits the model trained with synthesized speech much more than the one with human speech. This can be explained by the absence of noise diversity in the synthesized speech. Common Voice is a dataset composed of a lot of environmental noises, whereas the synthesized speech just has some artifacts, but not environmental noises. {Therefore, with the use of AA, the gap between models trained only with human and synthesized speech is reduced to a relative difference of 51.03\% and 55.86\% for those two languages}.

{Our results are interesting because, despite using only a single-speaker dataset for training the TTS model for pt-BR and ru-RU, our ASR model trained only with synthesized speech achieves a comparable result to related works. Considering \cite{rosenberg2019speech}, the relative difference between models trained only with human and synthesized speech was of 78.98\%; and ours around 60\% for two non-English languages.  Even though this work explores the English language in a setting with many available speakers and it is not directly comparable, we believe that our results indicate that our approach requiring just 1 speaker in the target language, can be used for low-resource languages.}

{In this way, we believe that the YourTTS model proposed previously meets the requirements of this paper, trained with just one 
speaker in the target languages is enough to be used for ASR data augmentation.}

\vspace{-0.2cm}
\section{{Is only one human speaker in the target language enough to get a reasonable ASR model via cross-lingual voice conversion and cross-lingual multi-speaker TTS ASR data augmentation?}}\label{sec:method:experiments-only-one}
\vspace{-0.1cm}

{In Section \ref{sec:method:generated_vs_human} we showed that YourTTS model trained with only a single speaker in pt-BR and ru-RU was effective as ASR data augmentation, {achieving similar results as previous works that explored multi-speaker datasets.}
{Although the TTS model was trained with only one target-language speaker, we have used all Common Voice dataset speakers' embeddings to create the synthetic version of this dataset. So, the speaker embeddings were extracted from target languages' native speakers.} 
This approach has shown good results, but many low-resources languages do not have datasets with many available speakers. So, the results reported are not realistic for extreme low-resource languages. For this reason, in this section we explored the use of a single-speaker in the target language, for training both TTS and ASR. To make up for the lack of speakers during the creation of the synthetic dataset, we have used English speakers, rather than cloning speakers from the target language.} {We created two synthetic datasets using YourTTS model:}

{\textbf{\textit{GEN\_TTS}}: Was created by synthesizing all the sentences in pt-BR and ru-RU Common Voice using English speaker {embeddings}, chosen at random from the 1,248 speakers available in the training set. {Differently from Section \ref{sec:method:generated_vs_human}, no target-language speaker embeddings are used because we focus on a more restricted, extremely low resource scenario. Each sentence was synthesized using one English speaker embedding.} We would like to note that, in preliminary experiments, we explored increasing the number of speakers per sentence from this dataset; however, this did not bring significant improvements. During the generation of this dataset, as explained previously in Section \ref{sec:method:yourtts}, diversity is achieved by randomly choosing $L$, $T$ and $T_{dp}$: for $L$, a value between 0.7 and 2, while for temperatures ($T$ and $T_{dp}$) a value between 0 and 0.667.}

{\textbf{\textit{GEN\_VC}}: consists of the single-speaker dataset used for training the TTS model in the target language converted to a multi-speaker dataset using cross-lingual voice conversion with English speakers. Each sample used in the TTS training was converted to the voice of 5 speakers, chosen at random from the 1,248 English speakers. The value of 5 transfers per sample was chosen in preliminary experiments. 
}

\begin{figure}[]
\centering
 \vspace{-0.1cm}
\resizebox{0.44\textwidth}{!}{%
\includegraphics[width=1\textwidth]{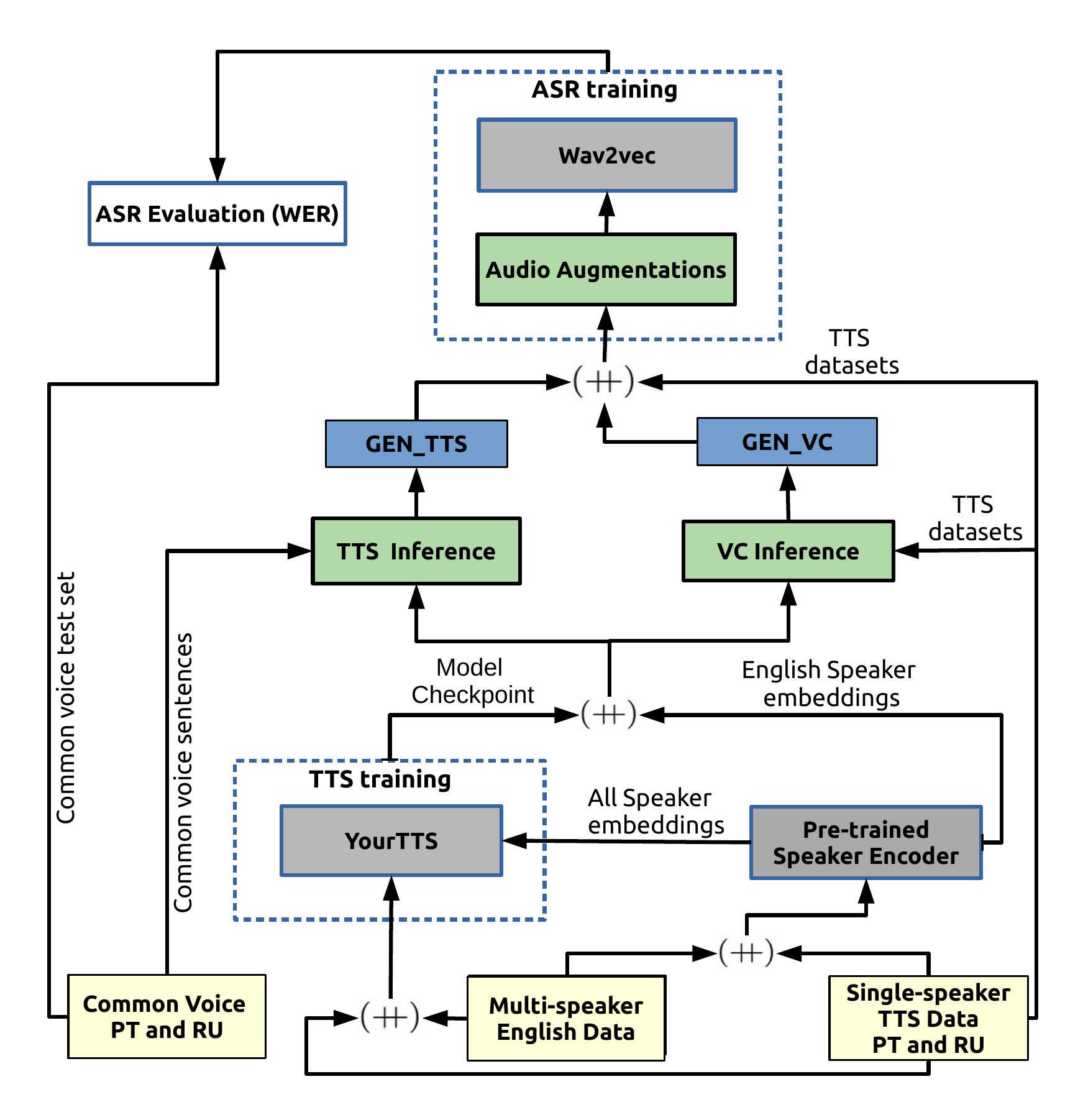}
}
 \vspace{-0.4cm}
 \caption{{Full pipeline diagram for Baseline + DA experiment}}
  \vspace{-0.6cm}
 \label{fig:generaltop}
\end{figure}

We carried out three experiments that used AA and the same training parameters used in Section \ref{sec:method:generated_vs_human}: 

\begin{itemize}[noitemsep]
    \item {\textbf{Baseline:} ASR models trained with the single-speaker dataset used during the TTS model training on pt-BR and ru-RU.}
     \item {\textbf{Upper Bound:} ASR models trained on pt-BR and ru-RU with Common Voice plus the single-speaker TTS dataset.}
    \item \textbf{Baseline + DA:} human speech from a single-speaker in the target language (TTS dataset), with data augmentation being accomplished by YourTTS. For data augmentation we merge the \textit{GEN\_TTS} and \textit{GEN\_VC} datasets, detailed above. {Figure \ref{fig:generaltop} presents a full pipeline diagram of this experiment.}
\end{itemize}

Table \ref{tab:results-comp2} presents our experiments' results on the {original} test subsets of the pt-BR and ru-RU Common Voice datasets.

\begin{table}[!ht]
\centering
 \vspace{-0.3cm}
\caption{Results on the pt-BR and ru-RU CV testsets (WER)}
 \vspace{-0.2cm}
\label{tab:results-comp2}
\resizebox{0.49\textwidth}{!}{%
\begin{tabular}{l|l|c|c}
\hline
\multicolumn{1}{c|}{\textbf{Experiment}} & \multicolumn{1}{c|}{\textbf{Train data}} & \textbf{PT}    & \textbf{RU}    \\ \hline
Baseline           &  TTS dataset (single-speaker)    &  63.90 &  74.02  \\ \hline 

 Upper Bound   &    Common Voice + TTS dataset          &  20.39 & 24.80  \\ \hline 
 Baseline + DA  &  TTS dataset +  \textit{GEN\_TTS} + \textit{GEN\_VC}  & 33.96 &  36.59  \\ \hline

\end{tabular}
}
 \vspace{-0.3cm}
\end{table}

{The model trained with just 1 target language speaker (Baseline) achieved a WER of 63.90\% and 74.02\% for pt-BR and ru-RU. The model trained with only 1 real target language speaker (TTS dataset) with data augmentation using voice conversion and speech synthesis (Baseline + DA), achieved a WER of 33.96\% and 36.59\%. Therefore, our data augmentation approach in scenarios with just 1 real speaker available improves the WER by 29.94\% and 37.43\% for pt-BR and ru-RU.}

{Comparing the results with the SOTA English ASR system on Common Voice (7.7\% achieved by \cite{zhang2022bigssl}), these results do not look so remarkable; however, \cite{quintanilha2020open} used {approximately}  158 {hours} of pt-BR speech and a non-self-supervised model without an external LM and achieved a WER of 47.41\% on the test set of BRSD v2 dataset. Despite using a different dataset, \cite{stefanel2021brazilian} showed that the Common Voice test set is more challenging than the test set of BRSD v2, and for this reason, the model proposed by these authors reached a higher WER on the Common Voice dataset.  In ru-RU, \cite{huang2020cross} used transfer learning from 5 large English datasets, trained the QuartzNet model \cite{kriman2020quartznet} on Common Voice, obtaining a WER of 32.20\% on the test set. Therefore, 33.96\% WER achieved by our model is probably superior to the SOTA for pt-BR, before the introduction of self-supervised learning approaches. Also, the WER of 36.59\% achieved in ru-RU is comparable with the SOTA.}

{Comparing the results of the Baseline + DA experiment with the Upper Bound (20.39\% and 24.80\% for pt-BR and ru-RU), our results still miss the Upper Bound. However, the results are remarkable since the TTS and ASR models were trained with just \textbf{one real speaker} in the target language, and the model was able to recognize the voice of over a thousand speakers from the Common voice test set.}

\vspace{-0.2cm}
\section{Conclusions and future work} \label{sec:conc}
\vspace{-0.1cm}

We presented a novel data augmentation approach for ASR training by using cross-lingual multi-speaker speech synthesis and voice conversion. We show that it is possible to achieve promising results for ASR model training with just a single-speaker dataset in a target language, making it viable for low-resource scenarios. Finally, our approach works both in a language (pt-BR) present in the Wav2Vec 2.0 {self-supervised} pre-training, as well as for a {completely} unseen language (ru-RU).  

{In future work, we intend to explore the use of a self-supervised model feature extractor as a discriminator during the training of the YourTTS model. In this way, the YourTTS model may produce even more human-like speech. In addition, we intend to do ablation studies using other SOTA ASR models like Whisper \cite{radford2022robust} and WavLM \cite{chen2022wavlm}. Finally, we intend to apply our method to Brazilian indigenous languages that have a few or even only one single-speaker data available.}

\section{Acknowledgements}
This study was funded by the Coordena\c{c}\~{a}o de Aperfei\c{c}oamento de Pessoal de N\'{i}vel Superior (CAPES) -- Finance Code 001, by CNPq (National Council of Technological and Scientific Development) grant 304266/2020-5, by  Artificial Intelligence Excellence Center (CEIA)\footnote{\url{http://centrodeia.org}}. The authors of this work also would like to thank the Center for Artificial Intelligence (C4AI-USP)\footnote{\url{https://c4ai.inova.usp.br/}} and the support from the São Paulo Research Foundation (FAPESP grant \#2019/07665-4) and from the IBM Corporation.

\clearpage
\bibliographystyle{RefStyle}

\bibliography{references}
\clearpage
\appendix

\section{Upper Bound + DA}

To verify how data augmentation using speech synthesis and voice conversion can improve the results of ASR models trained with multiples humans speakers, even when the TTS/voice conversion model was trained with only a single speaker in the target language. We train the ASR models with the merged datasets from \textit{GEN\_TTS}, \textit{GEN\_VC}, TTS dataset and Common Voice. We called this experiment as Upper Bound + DA. 

Table \ref{tab:results-comp-ap} presents results of the experiments Upper Bound (reported previously on Section \ref{sec:method:experiments-only-one}) and Upper Bound + DA experiments on the test subsets of the pt-BR and ru-RU Common Voice datasets.

\begin{table}[!ht]
\centering
\caption{WER of Upper Bound  and Upper Bound + DA experiments on the test subsets of the pt-BR and ru-RU Common Voice datasets}
\label{tab:results-comp-ap}
\resizebox{0.49\textwidth}{!}{%
\begin{tabular}{l|l|c|c}
\hline
\multicolumn{1}{c|}{\textbf{Experiment}} & \multicolumn{1}{c|}{\textbf{Train data}} & \textbf{PT}    & \textbf{RU}    \\ \hline
 Upper Bound   &    Common Voice + TTS dataset          &  20.39 & 24.80  \\ \hline
 Upper Bound + DA & \makecell{Common Voice + TTS dataset \\+ GEN\_TTS + GEN\_VC }& 20.20  &  19.46\\ \hline 

\end{tabular}
}
\end{table}

The model trained with Common Voice and a single-speaker TTS dataset in the target languages (Upper Bound) achieved a WER of 20.39\% and 24.80\% for pt-BR and ru-RU, respectively. The model trained with Common Voice, a single-speaker TTS dataset and our data augmentation approach in the target languages (Upper Bound + DA) achieved a WER of 20.20\% and 19.46\%, respectively.
{Therefore, the ASR model's trained with our data augmentation approach achieved a relative improvement of 0.93\% and 21.53\%, respectively, for those languages. The relative improvement achieved in pt-BR is consistent with the results reported in \cite{rosenberg2019speech}, where a relative improvement was of 0.79\%. However, it is lower than the result reported by \cite{li2018training} (4.56\%)  and \cite{laptev2020you} (4.25\%). On other hand, the relative improvement achieved in ru-RU is significantly superior than taht which was reported in related works.}

Unlike related works, we use AA and only one target-language speaker in the training of the TTS/voice conversion model. AA can make the ASR model more robust and improve generalization; however, these approaches may have overlapping contributions. To verify this hypothesis, we did an ablation study by re-training the Upper Bound and Upper Bound + DA experiments in pt-BR without AA. The Upper Bound achieved a WER of 22.96\% and the Upper Bound + DA achieved a WER of 21.41\%. {That is, without the use of AA as in the related works, the relative difference between ASR models trained only with human and only synthesized speech in pt-BR is 6.75\%.
Thus, in pt-BR and ru-RU, our approach achieves results better than related works in English, using only one speaker for training the TTS/voice conversion model in the target language. In this way, we have shown that it is possible to apply TTS systems to ASR dataset generation even for languages where just a single-speaker dataset is available. }

We believe that the difference between the results achieved for pt-BR and ru-RU, can be explained by the characteristics of the datasets. In Common Voice, the amount of hours available for training the ASR model for the ru-RU is 25.95 hrs as opposed to 14.52 hrs for pt-BR. Furthermore, the ru-RU TTS dataset has approximately 6.5 more hours after preprocessing than pt-BR and the ru-RU TTS dataset is high-quality. In our experiments, we use voice conversion to transform the TTS dataset into a multi-speaker dataset using 5 transfers for each sample, thus the difference in the number of hours is a multiple as well. 


\end{document}